# Using Edge Cases to Disentangle Fairness and Solidarity in AI Ethics
James Brusseau

### Abstract
Principles of fairness and solidarity in AI ethics regularly overlap, creating obscurity in practice: acting in accordance with one can appear indistinguishable from deciding according to the rules of the other. However, there exist irregular cases where the two concepts split, and so reveal their disparate meanings and uses. This paper explores two cases in AI medical ethics – one that is irregular and the other more conventional – to fully distinguish fairness and solidarity. Then the distinction is applied to the frequently cited COMPAS versus ProPublica dispute in judicial ethics. The application provides a broader model for settling contemporary and topical debates about fairness and solidarity. It also implies a deeper and disorienting truth about AI ethics principles and their justification.

## 1. Introduction
Contemporary artificial intelligence analyses of human skin lesions are producing an unusual combination of medical device efficacy, and patient vulnerability. The combination will be explored to disentangle two principles routinely knotted together in AI ethics: fairness and solidarity.

The disentangling is required precisely because it makes a practical difference so rarely. In most situations, the application of one principle leads to nearly the same outcome as applying the other, and the congruence creates a natural obliviousness to the two concepts' difference. Even the fact that there are distinct definitions easily goes unperceived. So, the initial difficulty with the split between fairness and solidarity is that it barely exists. In the case of artificial intelligence and skin cancer, however, it will exist, and lethally.

## 2. The principles of fairness and solidarity

Fairness traces back to Aristotle's definition as equals treated equally, and unequals treated proportionately unequally (Aristotle 1934: Book 5:3:13). Two comparable patients similarly threatened by cancer should receive parallel access to medical care and, by the same logic and with the same vigor, patients suffering dissimilar threats receive divergent care. Fairness imposes equality, and inequality.

Solidarity is inclusiveness and the demand that no one be left behind. Operationally, the principle distributes the maximum advantage to the most disadvantaged, to those who have least. The ideas of "justice" and "social justice" can adhere to this concept, joined more recently by the term "equity." Regardless, the underlying logic is inclusion: if the resources available to screen for skin cancer are limited, their distribution will be tilted toward those who normally have the least access to medical services.

There is a tipping point question in the ethics of solidarity: How much should the privileged sacrifice so that no one gets left behind? The American philosopher John Rawls responded with this principle: the *greatest* benefit must always go to the least advantaged (Rawls 1971: 302), which is different from all benefit. So, breakthroughs in medical AI will be distributed to benefit everyone, but mostly to benefit the underserved. It follows that Rawls does not automatically promote pure equality in the sense of everyone ultimately receiving exactly the same medical service, but he leans in that direction because the underserved are constantly receiving disproportionately more access.

A stronger imperative to inclusion emerges from French writer Simone de Beauvoir's *Who Shall Die?* (1983/1945) where she envisions a town under wartime siege, facing the rising dilemma of insufficient supplies to feed everyone. Instead of divvying out the remaining rations to at least some citizens, perhaps the hungriest or weakest, the town decides to collectively launch a suicidal attack on the enemy. At the extreme, solidarity means that if one must die, all do. No one left behind.

Beyond the definitions, what is significant about fairness and solidarity is that while the principles frequently overlap, they can also diverge, with the implication that fairness accompanies the injustice of failed solidarity, or solidarity coexists with unfairness. The distribution of a medical service by lottery, for example, is fair since all patients have an equal opportunity for treatment, but it fails solidarity since some are left behind, untreated.

## 3. Cases

Real experiences in AI ethics present more nuanced examples. The cases in this paper's background were produced by a team of philosophers, computer scientists, doctors, and lawyers organized out of the Frankfurt Big Data Lab in Germany (Brusseau 2020). The group approaches AI-intensive startup companies to collaboratively explore their experiences in technological development, and then reacts with a discussion and written report. Attention splits between ethics, technology, law and medicine (Zicari *et al* 2021a, Zicari *et al* 2021b).

Our skin lesion case started at the German Research Center for Artificial Intelligence (DFKI) and with a team led by Andreas Dengel. They developed an explainable AI in dermatology product – exAID – which translates machine learning outputs into human medical language. The underlying diagnosis is produced from skin lesion images processed by a neural network and converted into a probability that the lesion is cancerous. What exAID adds to the raw number is an explanation of which specific qualities in the image contributed to the AI finding. Along the development's way, an unusual, twin bias emerged. The image analyzing technology functions better on lighter skin, which immediately raises questions about data and algorithmic bias converting into racial discrimination. However, it is also true that skin cancer is more prevalent among the lighter skinned, so the imbalances are parallel: the AI discriminates, but the disease also discriminates, and in the same direction (Zicari et al. 2021b).

The second case is Danish and documents a more conventional arrangement of medical service distributed unevenly. A group led by Stig Nikolaj Blomberg (Blomberg et al. 2021) formed an AI startup to analyze frantic 112 calls – Denmark's 911 – for humanly imperceptible clues suggesting that the subject was suffering cardiac arrest. When perceived, the machine alerted a dispatcher to send a specially equipped ambulance. Because the essential process was natural language processing, disparities could be predicted: accents, dialects, and native language differences might affect diagnostic accuracy at the cost of linguistic minorities. Blomberg reported that Swedish and English speakers were well represented in the training data, alongside the native Danish, but in cosmopolitan Copenhagen, it is difficult to account for every verbal expression (Zicari et al. 2021a). The result is an unequal distribution of healthcare across a population of callers in equal jeopardy.

## 4. Fairness and solidarity and skin cancer

The skin cancer recognition technology is an edge case in fairness because it presents a bias that is offset by the same difference originally causing the disparity: the lighter skinned are more likely to be correctly diagnosed for a disease that infects them more frequently than the darker skinned. Counterintuitively, the AI is fair not despite the fact that it is more sensitive for one race than another, but *because* of the disparity. The imbalanced logic starts from the empirical inequality in melanoma: the lighter and darker skinned – whites and blacks in racial terms – are unequally susceptible to the lethal cancer. The corresponding ethical imperative is unequal treatment, which is received in the form of technological performance biased toward the lighter end of the skin-tone continuum. So, unequals are treated unequally, the definition of fairness.

In this edge case, technical bias is not an obstacle to ethical fairness, it *is* ethical fairness.

Going forward, there are decisions to be made. As our group noted in our report, any medical device that functions disparately across races or genders will encounter questions about whether resources should be reallocated to eliminate treatment differences. Strictly within the AI ethics of fairness, the response is no. Engineers should keep developing as they are, using available data to improve the diagnostic capacities of their digital tools along the established line of imbalanced treatment corresponding with uneven vulnerability. Sensitivity (the detection of malignancy) and specificity (the ratio of true alerts against false alerts) will improve along the entire skin-tone spectrum, without closing the racial quality gap.

Solidarity is different. The value of leaving no one behind closes the gap, and will do so even at the cost of healthcare treatment overall. The direct route is to devote the most resources to those who have least, the advantage to the disadvantaged. Engineers will need to reorient their efforts to improve diagnostic performance for the image analysis as applied to the under-served, the darker skinned. Theoretically, the unbalanced refining could continue until the diagnostic equipment comparably serves all those across the racial spectrum. The goal would be equal treatment for all potential patients. Here, however, the equality is unfair: unequals are treated equally in the sense that the more vulnerable and the less vulnerable are protected to the same degree by the AI.

As a practical matter, *pure* solidarity is impossible because in the real world a medical device must first work for some one person before service can expand to apply equally to everyone. Still, the ideal can be pictured as a thought experiment. Just as de Beauvoir's citizens agreed that if one must die, all must, so too in the case of AI analysis of dermatological images, the theoretical case could be made that until the machine works equally well for everyone, it should be used by no one.

Details are important. If fairness and solidarity demand specific and distinct allocations of resources, how can those disbursements be measured? Of what are they composed? What do the two principles actually mean for medicine and one or another skin-toned patient? Part of an answer involves training data: since whites are more likely to suffer the disease, it is unsurprising that more images have been taken of lighter skin and added to the image sets used to instruct pattern-recognition technology to discern cancer. This partially explains why today's machines function better on the light end of the skin spectrum (Zicari et al. 2021b). To the extent that training data correlates with model performance, it follows that a lever exists to adjust the fairness and solidarity balance. Crudely, the case could be made that resources should be expended to gather pictures of lighter-skinned lesions at a proportion greater than those gathered for the darker-skinned, and at the same rate of difference as the risk of the disease. Unequals would be treated unequally by counting pictures. Or, the data could be tilted toward solidarity by capturing more darker-skin images, perhaps to the point where all six of the scientifically delineated skin types on the Fitzgerald scale are equally represented in the picture set. This way, no one is left behind. Either way, the critical point is that being fair happens without solidarity, and solidarity is achieved without fairness.

## 5. Fairness and solidarity and cardiac arrest

The cardiac arrest recognition case also intersects with fairness and solidarity, but in terms of language and nationality, instead of skin and race. More significantly, fairness and solidarity overlap in Copenhagen instead of diverging.

The potential subjects of cardiac arrest at the center of the monitored phone calls are, as patients, equals: whether they speak Danish or Polish or Arabic is not material to their cardiac health. Unlike the case where skin color and skin cancer link intrinsically, here speech distinctions can neither justify nor preclude treatment differences. So, with the equality established as cardiac health irrespective of linguistic condition, the treatment must also disregard language, and that is a problem for Blomberg and his team if they want to be fair. As our group reported:

> Non-native speakers of Danish may not have the same outcome. Swedish and
> English speakers were well represented in the training data but the concern is
> that the training data may not have a diverse enough representation (Zicari et al.
> 2021a).

Equal patients receive unequal treatment from an artificial intelligence that helps
speakers of some languages more than others. International visitors and immigrant
communities are two potential sources of locally uncommon speech, but regardless
of the origin, the conclusion is that the cardiac arrest AI is unfair to comparable
patients when the words of some are well-recognized while others are mechanically
garbled.

Solidarity reaches the same conclusion, but by a different route. Instead of the logic
of equality, it is the imperative to inclusiveness that demands the natural language
processing be re-weighted to distribute the most to those who have least: if no
patients are going to be left behind, the AI must expand toward understanding the
languages of them all, even the rarest and most remote. Stronger, it is because of
rarity and remoteness that the speech's priority elevates. As we noted in our report:

> There is a risk that the AI system does not work equally well with all ethnic
> groups. It works best with Danish-speaking callers. It actually has a lower
> degree of being able to handle caller diversity than the dispatchers, who
> sometimes speak several languages. Thus, ethnic minorities would be
> discriminated against. (Zicari et al. 2021a)

What is most remarkable about the evidentiary blockquote just above is that it could
be switched with the blockquote in the preceding paragraph, and with no practical
difference.

The switch would pass unperceived because the overlap between fairness and
solidarity in this case – as in many cases – is nearly complete both in terms of what
triggers ethical attention (the technology serves some users better than others), and
in terms of a practical response (the technology's usefulness expanded to balance
performance across diverse populations). What should be underlined, though, is
neither the common trigger nor the common response, but instead the difference
hidden between the two similarities: the *reason* for linguistically expanding the AI
training. The solidarity reason is not based on equality, but on inclusiveness. And
that distinction remains, even while also remaining practically invisible. Two people
could work hand in hand to expand the machine's language perceptiveness, and
both be motivated by ethics without realizing their ethics were plural, even
unrelated. Sitting together, neither would realize their separation.

Ultimately, pragmatism trumped theoretical purity and the Copenhagen AI was optimized for the languages commonly heard there. In order to provide maximum help with limited resources, some people were treated unfairly while also suffering from a lack of solidarity, and all the while the difference between the two ethical concepts remained virtually imperceptible. What the skin lesion case showed, contrastingly, is that even though fairness and solidarity may frequently be a distinction without a difference, it is still a distinction, and it can make a difference.

### 6. Fairness and solidarity and COMPAS and ProPublica

Clear and distinct definitions of fairness and solidarity make a difference in the iconic AI ethics dispute (Bartneck et al. 2021) between the Northpointe software firm and the ProPublica social activist organization. Northpointe's Correctional Offender Management Profiling for Alternative Sanctions (COMPAS) algorithm produces a score between 1 and 10 predicting whether a defendant currently facing criminal charges will go on to be arrested for yet another crime if released from jail. The score helps judges determine whether defendants should be freed on bail pending trial.

To govern the predictions in a racially diverse environment, COMPAS employed classical fairness: white and black defendants who were equally likely to commit another crime if released were assigned an equal score. Those unequally likely to reoffend were assigned proportionately unequal numbers.

ProPublica advocated for an advantage to be distributed to the disadvantaged, and the strategy took form within an initial imbalance in the set of defendants. Blacks reoffend at a higher rate than whites, so relative to the population of those predicted to recidivate, a larger proportion of blacks than whites is correctly predicted to reoffend and so denied release from jail, and a larger proportion is incorrectly predicted to reoffend and so also denied release from jail. The split attracted ProPublica's attention toward one imbalance (larger percentage of blacks incorrectly jailed), and away from the other (larger percentage of blacks correctly jailed).

Concretely, distributing an advantage to the disadvantaged started by locating the disadvantaged. In the environment of defendants, that corresponded with the merciless reality of those jailed even though they would not have reoffended, and within that group, as ProPublica documented, blacks suffered harsher treatment from the courts than whites in the sense of being unnecessarily jailed at a higher rate. (The degree of disparity changes when measured between racial groups, or against the entire population of those needlessly held, but the fact of the disparity remains.) From there, ProPublica proposed that the number of black defendants needlessly jailed be limited to the lower proportion established by white defendants. Statistically and practically, there are a number of different calculations for implementation (Sahil and Rubin 2018), but conceptually, the adjustment is not difficult: black defendants predicted by the COMPAS algorithm to reoffend are freed until the number of false positives (defendants predicted to reoffend who do not) drops to match the level set by whites. The result is inclusion in the sense that among the worst off, no one racial group suffers disproportionately. (See *Figure 1*.)

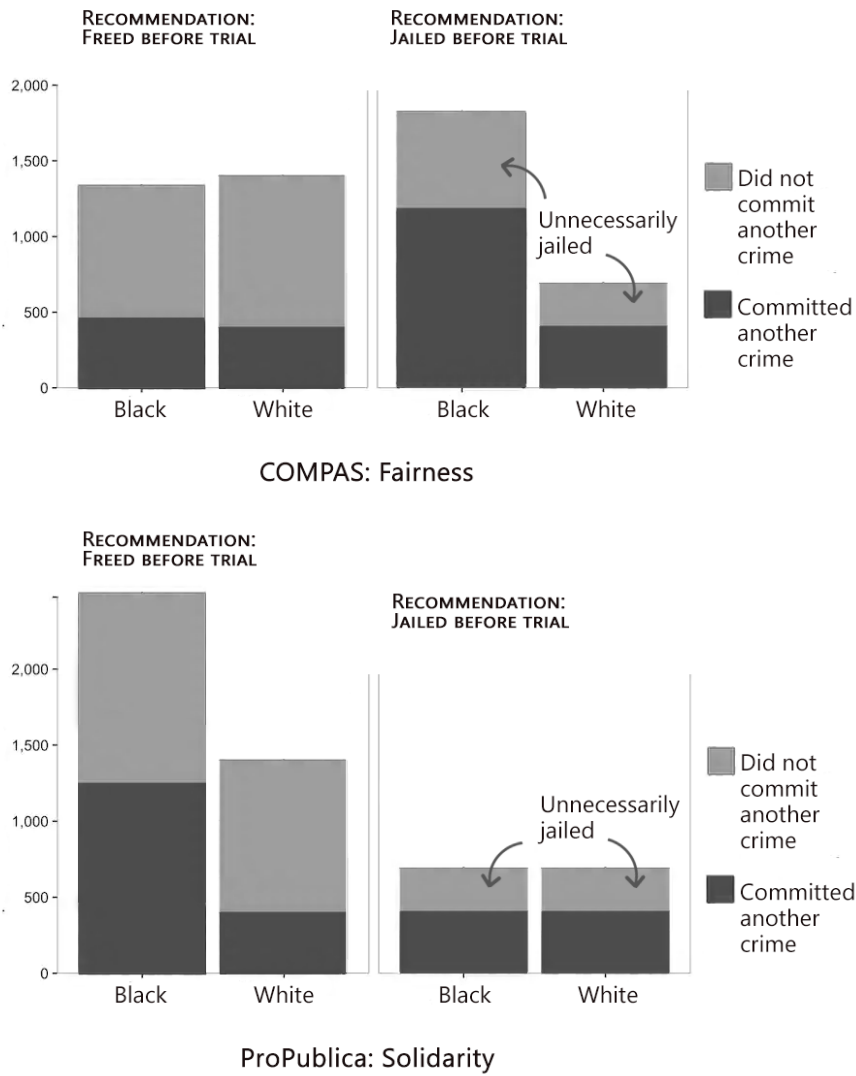

**Figure 1.** These tables are conceptually correct, though the specific numbers are in dispute (Barenstein 2019). The specific numbers are provided to present a sense of the scale of the ProPublica study, and its general results.

If implemented, the ProPublica strategy projects two results: release for many black defendants whose criminal profiles match imprisoned whites, and balanced suffering for those blacks and whites unnecessarily imprisoned.

This is blatantly unfair, but also exemplary solidarity. It is unfair because equals – black and white defendants with the same criminal profile – are treated unequally. It is solidarity because the entire system is attuned to those who are disadvantaged, and to provide that group with an advantage, which is the minimization of unnecessary imprisonment down to the point where no one group is left behind. This solidarity is the true goal of ProPublica, as they state in one of their written exchanges with the COMPAS providers: "The whole point of due process is accuracy to prevent people from being falsely accused. (Angwin and Larson 2016)"

Orienting justice to advantage the disadvantaged by preventing false accusations and unnecessary imprisonment is a noble goal, but it has nothing to do with fairness. With respect to fairness, there is *no difference* between optimizing so that no one who is innocent is imprisoned, and optimizing so that no one who is guilty goes free. All that matters is that the rules for deciding who goes where are equally applied. For solidarity, by contrast, equal application – and the logic of equality itself – is either irrelevant or counterproductive. What matters is only that those who are in the worst position receive the most help.

Finally, resolving the Northpointe and ProPublica dispute requires two inseparable sentences: Northpointe advocates fairness in incarceration ethics, and ProPublica advocates solidarity. Between the two, there is no intrinsically right or wrong choice.

### 7. Fairness and solidarity untangled
It is definitely wrong, however, to imagine that Northpointe and ProPublica are involved in an ethics *debate*. Debating implies sharing a set of common presuppositions on which disagreements are presented, and there is no common ground here: what counts as ethics, what the word means, is localized distinctly on each side. The reasons – and the *kind* of reason – guiding decisions about recidivism and incarceration are literally incomprehensible from one to the other.

It is common in AI ethics today to find the COMPAS and ProPublica approaches to recidivism characterized as "incompatible" (Rudin et al. 2020), or as "mutually exclusive" (Hao and Stray 2019) or as "competing and incompatible" (Rahwan et al. 2019). These descriptions are only slightly imprecise, but the imprecision's effects are expansive. Instead of incompatible, the two ways of doing ethics are *incomparable*. They are not like rain and snow, they are like rain and Mondays. Sometimes you have rain, sometimes you have Monday, sometimes you have both overlapping, and other times neither. In AI ethics, we frequently have fairness and solidarity together, but not because the two concepts are similar, instead, only because they overlap: they both function in about the same way at about the same time and with about the same material. This describes the cardiac arrest case in Denmark. In other cases, we have fairness without solidarity, which occurred in the skin lesion example, and in the COMPAS algorithm. Still other times, we have solidarity without fairness, which occurred in the ProPublica objection to COMPAS. In any case, and in *every* case, what must be escaped is the presupposition that either fairness or solidarity includes or excludes the other. They are neither compatible nor incompatible because they are neither the same nor different.

When that presumption of incompatibility is escaped, fairness and solidarity come untangled. They may be discussed clearly because they are understood entirely distinctly. This clarity and distinction is beneficial as it provides neat resolutions to disputes in AI ethics. Like all truths, however, it is also dangerous because there is no way to limit what it reveals.

## 8. Conclusion

The Northpointe COMPAS versus ProPublica dispute is not a problem, it is a solution. It solves the problem of fairness and solidarity tangled together in AI ethics by presenting a case where it is impossible to reconcile the principles, and also impossible to use one to exclude the other. Like the skin lesion case, the paradox of two concepts that are neither the same nor different ultimately forces a string of positive understandings. First there are the independent understandings of what fairness and solidarity mean, and then there comes the understanding *about* understanding, which is that no adjudication exists for them: there is no way to privilege one or the other with reasoning that appeals to both sides because they respond to disparate logics, to equality, or to inclusiveness. Between, there is only the reality that neither one makes sense to the other.

Tolerating this incomprehension will be unnatural for mainstream AI ethics, partially because of the composition of contemporary participants. The collection of those doing AI ethics today resembles our members organized from the Frankfurt Big Data Lab in weighing heavily toward computer scientists. Having worked with them, one observation is not surprising: their ethical methods resemble machine learning practices. In machine learning, knowledge starts as accuracy, like a prediction of cardiac arrest generated from speech patterns. Then, it advances by increasing breadth and depth. This is a wider data pool, more words, dialects and languages recognized and incorporated into ever deeper understanding registered as the algorithmic prediction's narrowing accuracy. Transferred from computer science to ethics, this funneling model for truth frequently works. It does in the cardiac arrest case where linguistic breadth and predictive accuracy swirl together with fairness and solidarity: more emergency telephone calls recorded and added to the machine learning for training equals increased accuracy, equals more fairness, equals more solidarity. Ideally, and at the end of knowledge and ethics, there is a machine that recognizes the words of all languages, and unerringly identifies every cardiac arrest, and for those reasons it is completely fair, and entirely just. We may never get there, but that is the destiny. Because more learning equals more knowing – just as more data equals more accuracy – the structure of learning itself is built to eliminate areas of incomprehension. The elimination is what learning *does*.

This explains why so many papers have been written in academic journals and presented to computer science conferences about COMPAS and ProPublica. The logic of the debate frustrates the methods of the debaters. Because the dispute is incomprehensible by nature, every attempt to extend broadly and gather the debate's ethical approaches into a single understanding can only fail, and therefore require still more expansive study allowing still deeper investigating. Already, computer scientists have broadened their considerations to include an eye-popping twenty-one separate definitions of fairness in the attempt to account for everything circulating through fairness and solidarity ethics, and even that is not enough, more definitions are being added (Verma and Rubin. 2018, Narayanan 2018: 00:54:00). Whether this circle is virtuous or vicious is not clear, but it is endless.

Out on the edges of ethics and artificial intelligence there is a different experience. Fairness and solidarity diverge, and getting closer to one means getting further from the other. Understanding one confuses the other. Ultimately, the splitting culminates and the two principles are no longer comparable, which is beneficial because the separation finally allows each one to be understood on its own terms, and without interference from the other. Capturing those divergent understandings was the purpose of this paper. It is not the end, though, because rendering fairness and solidarity as incomparable also implies a disorientation that will need to be managed elsewhere: it is impossible to explain *why* one should be privileged over the other. Those forced to decide between them will have to judge not despite imperfect justification, but because of the impossibility of any justification at all.

### Declaration

Author declares that there is no conflict of interest or external funding.

(Other standard declarations non-applicable.)

**END**